\documentclass[10pt,conference,letterpaper]{IEEEtran}
\IEEEoverridecommandlockouts
\usepackage[T1]{fontenc} 
\usepackage{cite}
\usepackage{amsmath,amssymb,amsfonts}
\usepackage{algorithmic}
\usepackage{graphicx}
\usepackage{tabularx}
\usepackage{textcomp}
\usepackage{xcolor}
\usepackage[hyphens]{url}
\usepackage{listings}
\usepackage{caption}
\usepackage{times}
\usepackage{multirow}
\usepackage{tikz}
\usepackage{hyperref}
\usepackage{lipsum}
\graphicspath{{figures/}}

\def\BibTeX{{\rm B\kern-.05em{\sc i\kern-.025em b}\kern-.08em
    T\kern-.1667em\lower.7ex\hbox{E}\kern-.125emX}}
\def\Csharp{C\#}

\makeatletter
\def\lst@makecaption{%
  \def\@captype{table}%
  \@makecaption
}
\makeatother

\newcommand\copyrighttext{%
  \footnotesize \textcopyright 2018 IEEE. Personal use of this material is permitted.
  Permission from IEEE must be obtained for all other uses, in any current or future 
  media, including reprinting/republishing this material for advertising or promotional 
  purposes, creating new collective works, for resale or redistribution to servers or 
  lists, or reuse of any copyrighted component of this work in other works. 
  DOI: \href{https://ieeexplore.ieee.org/document/8605772}{10.1109/UCC-Companion.2018.00049}}
\newcommand\copyrightnotice{%
\begin{tikzpicture}[remember picture,overlay]
\node[anchor=south,yshift=10pt] at (current page.south) {\fbox{\parbox{\dimexpr\textwidth-\fboxsep-\fboxrule\relax}{\copyrighttext}}};
\end{tikzpicture}%
}

\begin{document}

\title{Comparison of FaaS Orchestration Systems
\thanks{
This work has been partially supported by the Spanish government through project ``Software Defined Edge Clouds'' (TIN2016-77836-C2-1-R) and by~the AWS Cloud Credits for Research program.
}
}

\author{\IEEEauthorblockN{Pedro Garc\'{i}a L\'{o}pez, Marc S\'{a}nchez-Artigas, Gerard Par\'{i}s, Daniel Barcelona Pons, \'{A}lvaro Ruiz Ollobarren \\ and David Arroyo Pinto}
\IEEEauthorblockA{\textit{Computer Engineering and Mathematics Department} \\
\textit{Universitat Rovira i Virgili}\\
Tarragona, Spain \\
Email: \{pedro.garcia, marc.sanchez, gerard.paris, daniel.barcelona, alvaro.ruiz\}@urv.cat,  \\ david.arroyop@estudiants.urv.cat}
}

\maketitle

\copyrightnotice

\begin{abstract}

Since the appearance of Amazon Lambda in 2014, all major cloud providers have embraced the ``\textit{Function as a Service}'' (FaaS) model, because of its enormous potential for a wide variety of applications.  As expected (and also desired), the competition is fierce in the serverless world, and includes aspects such as the run-time support for the orchestration of serverless functions.  In this regard,  the three major production services are currently  \textit{Amazon Step Functions} (December $2016$), \textit{Azure Durable Functions} (June $2017$), and \textit{IBM Composer} (October $2017$), still young and experimental projects with a long way ahead.
In this article, we will compare and analyze these three serverless orchestration systems under a common evaluation framework. We will study their architectures,  programming  and billing models, and their effective support for parallel execution, among others. Through a series of experiments, we  will also evaluate the run-time overhead of the different infrastructures for different types of workflows.

\end{abstract}


\section{Introduction}

Serverless computing, and in particular, the ``Function as a Service'' (FaaS) paradigm, has taken the industry by storm.~In the last four years, all the major cloud providers have offered their own solutions, including the ``Big Four'' cloud vendors, namely,  AWS Lambda, IBM Cloud Functions, Google Cloud Functions, and Azure Functions.  Indeed, their ability to enable event-driven computing and scale to thousands of concurrent functions have spurred many cloud users to adopt serverless computing for a variety of applications, such as microservices, IoT, machine learning inference, etc. 

Unfortunately, the FaaS model is very young and it lacks of adequate coordination mechanisms between functions. Simply put, it is still \emph{cumbersome} to orchestrate  a pool of  serverless functions to build a complex application. Proof of that is that in different domains such as enterprise workflows, Web mashups, genomics pipelines, or even AI workflows, it is still complex to create flexible data processing pipelines and ensembles using serverless functions.

In this piece of research, we will compare the three main production services for function orchestration: \emph{IBM Composer}, \emph{Amazon Step Functions},   and \emph{Azure Durable Functions}. To the very best of our knowledge, this is the first work that presents a comprehensive comparison of  function orchestration systems.

\medskip
\noindent\textbf{Contributions.} The contributions are the following:
\begin{itemize}

\item \emph{Evaluation framework: }A first contribution is a set of tests and metrics for the evaluation of  serverless orchestration systems.  Our evaluation criteria is wide ranging, 
from the analysis of  their architectures, programming  and billing models, to  their effective support for massive parallelism. Furthermore, it includes a set of benchmarks to gauge the orchestration overheads in different scenarios.

\item \emph{Comparison of three major vendor  models}: By means of the aforementioned evaluation framework, our second contribution is a rigorous comparison of three commercial projects. We present interesting insights and results of the evaluation of the three systems for different applications.

\item \emph{Novel Suspend API:} As a third contribution, we propose a novel programming abstraction that will facilitate the implementation of custom orchestrators that comply with the \textit{serverless trilemma}~\cite{trilemma},
a recent lemma that identifies three  key competing constraints for function composition.

\end{itemize}

\section{Related Work}

The first related work is~\cite{trilemma} by Baldini et al. from IBM, which introduces the~\textit{serverless trilemma}: Functions should be considered as\emph{ black boxes}; function composition should obey a \emph{substitution principle} with respect to synchronous invocation (i.e., a composition should be also a function); and invocations should not be \emph{double-billed}. Also, the authors advocate for run-time support for function orchestration, and present  a solution for sequential compositions
that fulfills the trilemma:  \textit{IBM Sequences}.

Other systems have tried before to orchestrate functions with no explicit run-time support. These solutions can be classified into two types: (I) functions to orchestrate  functions; and~(II) external client schedulers. 
In the first category (e.g., \cite{flint, MALAWSKI2017}), the orchestration is performed inside a serverless function. However, this approach suffers \textit{double billing} according to the trilemma: The orchestrator function is billed while waiting for the execution of the orchestrated functions to complete (which are also billed).  In the second type (e.g., ~\cite{PyWren2017, ttt}), an external client scheduler coordinates functions, thereby avoiding double billing. But in this case, the \textit{substitution principle} is violated. Compositions cannot be treated themselves as functions since they are external to the  system itself. 

As we will see in the next section, our evaluation framework goes beyond the trilemma since it establishes different metrics to compare orchestration mechanisms.  Of course, we will also include the trilemma in our framework  as a  metric.

\section{Evaluation Framework}

To evaluate different orchestration services, we will consider the following metrics:

\begin{itemize}
\item \emph{ST-safeness:} One key aspect of any orchestration service is whether or not it fulfills the \textit{serverless trilemma}~\cite{trilemma} (ST) mentioned in the previous section. 
An orchestration service that complies with the trilemma is said to be \emph{ST-safe}.
\item \emph{Programming model:}  It refers to programming simplicity and the set of  coding abstractions, but also, to whether it
provides a \textit{reflective API} to observe the current state of~a function composition. 
\item \emph{Parallel execution support:} Whether the framework supports the orchestration of \textit{parallel} functions. 
\item \emph{State management:} How data is passed from one stage of a function composition to the next.
\item \emph{Software packaging and repositories:} Modularization and software reuse of serverless applications.
\item \emph{Architecture}: The orchestrator can be an external entity not implemented as a function (\emph{client-side scheduler}), or as part of the run-time itself  as a function, scheduled in reaction to events. 
For brevity, we will often refer to the latter  with the term ``\emph{reactive core}''~\cite{trilemma}.
 \item \emph{Overhead:} Given the reliance of orchestration services on a function scheduler,  the significance of the  orchestration overhead should be measured for 
representative function compositions such as chains and parallel patterns. 
\item \emph{Billing model}: To complete the picture, it is  fundamental to  provide detailed accounting, so users understand how much they need to pay.
\end{itemize}


Although the above list of metrics could be expanded, we believe that is by far enough to analyze the quality of the  three commercial projects.
A summary of the complete comparison can be found in Table~\ref{tab1}. Next, we evaluate  each project based on the above criteria, 
except the overheads that are reported~in Section~\ref{sec:val}.

\subsection{Amazon Step Functions (ASF)}

Amazon released ASF in December $2016$ with the aim to harness the composition of serverless functions. In particular, ASF allows the creation of workflows as finite state machines written in Amazon States Language, a custom JSON-based Domain Specific Language (DSL).


\medskip
\textit{ST-safeness.}  First of all, ASF does not comply with the serverless trilemma (i.e., not \emph{ST-safe}) because it breaks the \textit{substitution principle}. That is, a composition of functions is not  a function. Step Functions can be invoked, receive a JSON input and generate a JSON output, they can orchestrate other functions, but they are not functions themselves. This means that a state machine cannot be part of another state machine.

\textit{Programming model.} Amazon States Language supports function chaining and branching (\texttt{if} statements), function retries, and parallel executions.  However, the DSL only permits the representation of \emph{static} graphs, and it is difficult to program for relatively complex workflows. 

It provides a basic reflective API to query the running state or to cancel the entire workflow.  Further, it offers monitoring capabilities thanks to the logs accessible using CloudWatch.

\textit{Parallel execution support.} ASF offers support for parallel programming workflows in the DSL. They allow up to $1,000$ state transitions per second with burst capacity of $5,000$ state transitions (per account per region). 

\textit{State management.} ASF restricts state passing between functions to only $32$KB. Since this information must be logged for fault tolerance in long-running workflows, this limit helps to presumably reduce the underlying storage overheads. 

\textit{Software packaging and repositories.}  To quickly deploy  sample or complete serverless applications,  Amazon offers the AWS Serverless Application Repository \cite{awsrepo}. Each application is packaged using the standard AWS Serverless Application Model (SAM) \cite{awssam}. An important limitation here is that SAMs cannot include Step Functions,  thereby disabling composite applications orchestrated by ASF.

\textit{Architecture.} It is based on an external client scheduler that synchronously interacts with the functions involved in the state transitions (Steps),  and logs each action to persistently record it. Each transition will recover the previous state from the log and run the next state in the workflow.  Again, we insist that the scheduler itself is not a function in the platform.

\textit{Billing model.} Amazon provides a clear billing model. As of July 2018 it charges $0.025$ USD per $1,000$ state transitions.

\subsection{IBM Composer}


IBM released an early solution to function composition in November $2016$ with \textit{action sequences} (IBM Sequences), a simple mechanism~to chain together a sequence of functions. 
In addition to the support for different languages, a very interesting property is that a sequence itself can be invoked as a function in another composition. 
Because IBM Sequences is built into the \textit{reactive core} of OpenWhisk (the substrate of IBM Cloud Functions), there is no double billing for the orchestration.

However, IBM Sequences only supports simple chaining of functions. To address that,  IBM released Composer~\cite{ibmcomposer} as a Tech Preview in October $2017$. Specifically, IBM Composer adds other composition patterns beyond sequences like conditional constructs, loops, retries, etc. 

\medskip
\textit{ST-safeness.}  A key difference with ASF is that from the beginning, all these function orchestration services were given run-time support in the reactive core, 
fulfilling the \emph{substitution principle} for the synchronous orchestration of functions. As a result, IBM can properly claim to be the first to implement  a  \emph{ST-safe} serverless run-time.


\textit{Programming model.}  IBM Composer provides a complete composition library in JavaScript with functions such as \texttt{composer.sequence}, \texttt{composer.if} or \texttt{composer.try}, among others, which synthesize OpenWhisk \textit{conductor actions} to implement compositions. Moreover,  it also includes several command line interface (CLI) tools, alongside a visual workflow interface  for compositions (IBM Cloud Functions Shell). The programming model is much simpler than Amazon's DSL. Altough it does not support parallel execution patterns, it offers simple CLI commands to expose functions as Web frontends, and to compose functions with external Web microservices.

Unfortunately, it does not provide a reflective API to control conductor actions, only visual monitoring of the platform logs using Kibana or the IBM Cloud Functions  Shell.

\textit{Parallel execution support.} IBM Composer does not currently support parallel execution of functions in a composition.

\textit{State management.} IBM allows $5$MB of state parameters passed between functions in orchestrations. As we will see in Section~\ref{sec:val}, however,  the overheads related to state passing can significantly grow with increasing parameter sizes.

\textit{Software packaging and repositories.}  IBM offers the so-called OpenWhisk packages \cite{ibmpac} to bundle together  functions and their triggers. Furthermore, it permits to publish and search packages in a  public namespace in the IBM Cloud. However, it is also true that the Amazon SAM standard and metadata is more detailed and powerful than that of IBM. SAM metadata model supports many parameters for serverless applications,  like memory allocations, timeout, resource dependencies, and events and triggers.

\textit{Architecture.} The software architecture of the orchestrator service is integrated in the \textit{reactive core}. In~\cite{trilemma}, it is described how this can be accomplished with the help of the so-called ``active ack'' mechanism, inspired in a pipeline bypass strategy. The idea is to bypass the system of records and to use directly message queues to forward results to the orchestrator (termed ``controller'' here). It is claimed in~\cite{trilemma} that
such an  event-based  controller  hugely reduces the  overhead of transitioning from one function to another (at least inside the OpenWhisk framework). We will verify whether this is true in Section~\ref{sec:val}.

\textit{Billing model.} IBM Composer is still in Tech Preview, so its billing costs remain opaque. IBM Sequences charge users only for all the function invocations that occurred as part of  the sequential composition.

\subsection{Azure Durable Functions (ADF)}

ADF is an experimental project that Microsoft published in June $2017$. It is probably the most ambitious orchestration service thanks to its advanced programming abstractions.

\medskip
\textit{ST-safeness.} According to the trilemma, ADF is \textit{ST-safe}. It complies with the composition as function constraint.

\textit{Programming model.} It has better programmability than the other two projects because they define workflows directly in C\# code. Using the powerful \texttt{async/await} constructs, it becomes easy to build stateful durable workflows. Specifically, the programming model supports function chaining, retries, parallel spawning (fan-out/fan-in), and the interaction with external asynchronous Web services. 

~It provides a complete reflective API that permits not only accessing the current state of a given orchestration, but also triggering events to an awaiting orchestration instance. It even advertises novel services like eternal orchestrations, persistent addressing with singleton orchestrations, or versioning.

\textit{Parallel execution support.} ADF provides the \textit{fan-out/fan-in} pattern to allow executing multiple functions concurrently and perform aggregations on the results.

\textit{State management.} ADF does not restrict the size of state parameters passed across functions. Because this information is logged for fault tolerance in long-running workflows, ADF stores the parameters larger than $60$KB in compressed form to avoid overhead penalties and reduce storage costs.

\textit{Software packaging and repositories.} Regarding software packaging, Microsoft provides a very simple packaging format \cite{azurezip} for deploying functions. It is also possible to export to other Microsoft software packaging standards like .NET assemblies and Microsoft Web Deploy for  Web packages. 

\textit{Architecture.} Its software architecture is an extension of the reactive core.  Specifically, the architecture is based on the \textit{Durable Task Framework},  which enables development of long-running workflows using a pattern called \textit{event sourcing}. This pattern stores all events produced by function calls and enables the event replay to restore a previous state. Events are stored using Azure Storage queues, tables and blobs to manage state and events.

The key benefit of this approach is to support long-running workflows where the durable function can be hibernated, and later restored, using event sourcing. This also means that all the orchestration function code must be deterministic.

\textit{Billing model.} ADF is also in Tech Preview. Hence, there is not a clear billing model. The project's web site suggests that users can be billed by the execution time of the  Durable Functions in the composition. Users may also be charged with unpredictable storage costs originated by event sourcing.  This is worrying, and the web site even suggests that depending on the code of the function, storage costs could become large.

\section{Experimental results}
\label{sec:val}

We evaluate the run-time overhead of Amazon's, IBM's and Microsoft's orchestration services. We consider as \emph{overhead} all the time spent outside the functions being composed, which~is easy to measure
in all platforms. For a sequential composition $g$ of $n$ functions  $g = f_1  \circ f_2 \circ \dots \circ  f_n$, it is just: 
\[
\small
\mbox{overhead }({g})=  \mbox{exec\_time}( g) -   \sum^n _{i = 1} \mbox{exec\_time}( f_i).
\]

It is important to note here that our overhead definition includes the delays between function invocations, and the execution time of the orchestration function (for IBM Composer and ADF) or the delays between state transitions (for ASF).

For all the tests, we listed only the results when 
functions were in \textit{warm} state. This implies that we
did not consider the cold start of spawning the 
function containers and VMs. Our focus here was on measuring the
overheads of running function compositions.
All the tests were repeated $10$ times.
Measurements were done during June and July of 2018.

~Functions were coded in Java in all platforms. The single exception was ADF, which does not currently support Java, but \Csharp. The orchestration functions were implemented in the default language available in each platform: Node.js for IBM Composer, and \Csharp~for ADF. ASF orchestration was specified in Amazon States Language with the aid of AWS Java SDK. IBM Sequences were statically defined by a command-line argument at deployment time.

\subsection{Sequences}

First, we quantify the overhead for sequential compositions of lengths $n \in \left\{5, 10,  20 , 40 , 80\right\}$ for the following services:
 IBM Cloud (Sequences and Composer); AWS Step Functions; and Azure Durable Functions. For simplicity, all the functions in the sequence were the same: A function that slept for $1$s, and then returned.
Listings~\ref{lst:composerSeq} and \ref{lst:adfSeq} show the implementation of the orchestration function for IBM Composer and ADF, respectively. Listing~\ref{lst:asfSeq} is the Java code used to generate the JSON-based equivalent state machine for ASF.

\begin{lstlisting}[caption=IBM Composer code for the sequences experiment.,
  label=lst:composerSeq,]
composer.repeat(40, 'sleepAction')
\end{lstlisting}

\begin{lstlisting}[caption=ADF code for the sequences experiment.,
  label=lst:adfSeq,]
for (int i = 0; i < NSTEPS; i++) {
  await context.
    CallActivityAsync("sleepAction", null);
}
\end{lstlisting}

\begin{lstlisting}[caption=Code that generates the JSON-based state machine for the sequences experiment using the AWS Java SDK.,
  label=lst:asfSeq,]
StateMachine.Builder stateMachineBuilder = 
  stateMachine()
  .comment("A Sequence state machine")
  .startAt("1");
for (int i = 1; i <= NSTEPS; i++) {
  stateMachineBuilder.state(String.valueOf(i), 
    taskState().resource(arnTask)
      .transition((i != NSTEPS) ? 
        next(String.valueOf(i + 1)) : end()));
}
StateMachine stateMachine = 
  stateMachineBuilder.build();
\end{lstlisting}

\noindent\textbf{Results}. The results for $80$ functions are not available for IBM Sequences and Composer because IBM Cloud has a limit of $50$ actions in any composition.
ADF was tested twice with \texttt{extended sessions} enabled and disabled. This feature allows the platform to hold orchestrator function instances longer  in memory, avoiding the default aggressive replay behavior of event sourcing.

\begin{figure}[t]
  \centerline{\includegraphics[width=0.5\textwidth]{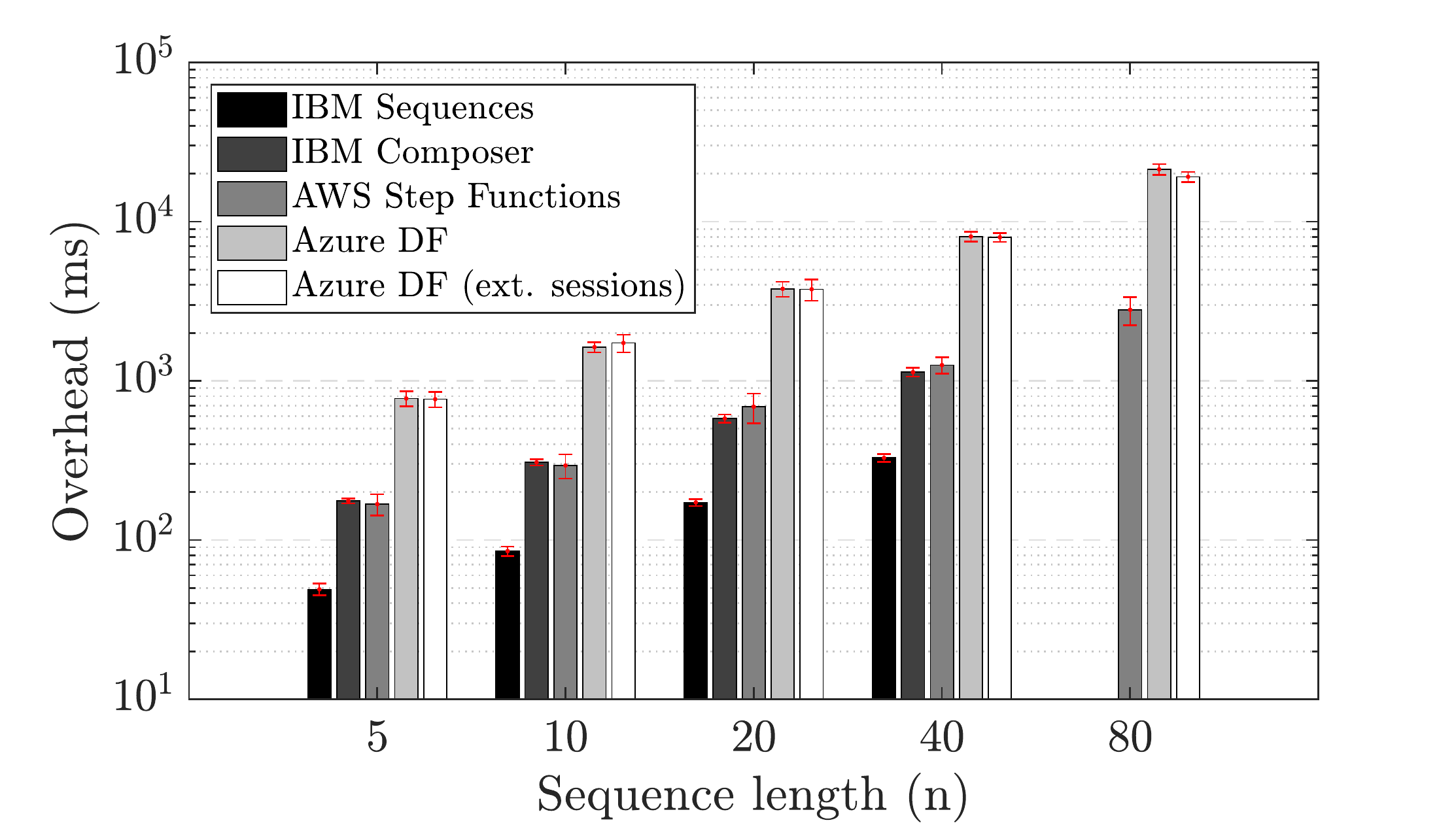}}
  \caption{Function sequences overhead.}
  \label{fig:seq}
\end{figure}

Fig.~\ref{fig:seq} plots the results for the different platforms. We see that the static compositions of IBM Sequences have the lowest overhead (around $0.3$s for $40$ functions). IBM Composer and AWS Step Functions exhibit a similar overhead ($1.1$s and $1.2$s, respectively, for $40$ functions). In comparison, Azure Durable Functions has a remarkably higher overhead (around $8$ seconds for $40$ functions), which does not significantly improve when using \texttt{extended sessions}.
Overall, the overhead grows linearly with the number of functions in the sequence.

Apparently, the limit of  $50$ actions in any composition, along with the lack of a waiting mechanism to suspend and resume orchestration at later times, disqualifies IBM Composer from implementing long-running workflows. However, ASF with its \texttt{Wait} state, and ADF with its \textit{durable timers}, are able to define long-running workflows that last for days or even months.

\subsection{Parallelism}

Our goal was to measure the overhead of running $n$ times the same function in parallel, for  $n \in \left\{5, 10,  20 , 40 , 80\right\}$.
The function  slept for $20$s and returned. So ideally, a zero-overhead parallel composition should last $20$s,  irrespective of the value of $n$.  The extra time was
pure overhead.  This experiment was only conducted for ASF and ADF (\texttt{extended sessions} enabled) --- IBM Composer does not currently support parallel execution. Listing~\ref{lst:adfPar} shows the ADF code for this test that uses the simple\textit{ fan-out/fan-in} pattern to execute multiple functions concurrently. Listing~\ref{lst:asfPar} does the equivalent for Step Functions.

\begin{lstlisting}[caption=ADF code for the parallelism experiment.,
  label=lst:adfPar,]
var tasks = new Task<long>[NSTEPS];
for (int i = 0; i < NSTEPS; i++)
{
  tasks[i] = context.CallActivityAsync<long>(
            "sleepAction");
}
await Task.WhenAll(tasks);
\end{lstlisting}

\begin{lstlisting}[caption=Code that generates the JSON-based state machine for the parallelism experiment using the AWS Java SDK.,
  label=lst:asfPar,]
StateMachine.Builder stateMachineBuilder = 
  stateMachine()
  .comment("A state machine with par. states.")
  .startAt("Parallel");

Branch.Builder[] branchBuilders = 
  new Branch.Builder[NSTEPS];

for (int i = 0; i < NSTEPS; i++) {
  branchBuilders[i] = branch()
    .startAt(String.valueOf(i + 1))
    .state(String.valueOf(i + 1), 
      taskState()
      .resource(arnTask).transition(end()));
}

stateMachineBuilder.state("Parallel", 
  parallelState().branches(branchBuilders)
  .transition(end()));
final StateMachine stateMachine = 
  stateMachineBuilder.build();
\end{lstlisting}


\begin{figure}[t]
  \centerline{\includegraphics[width=0.5\textwidth]{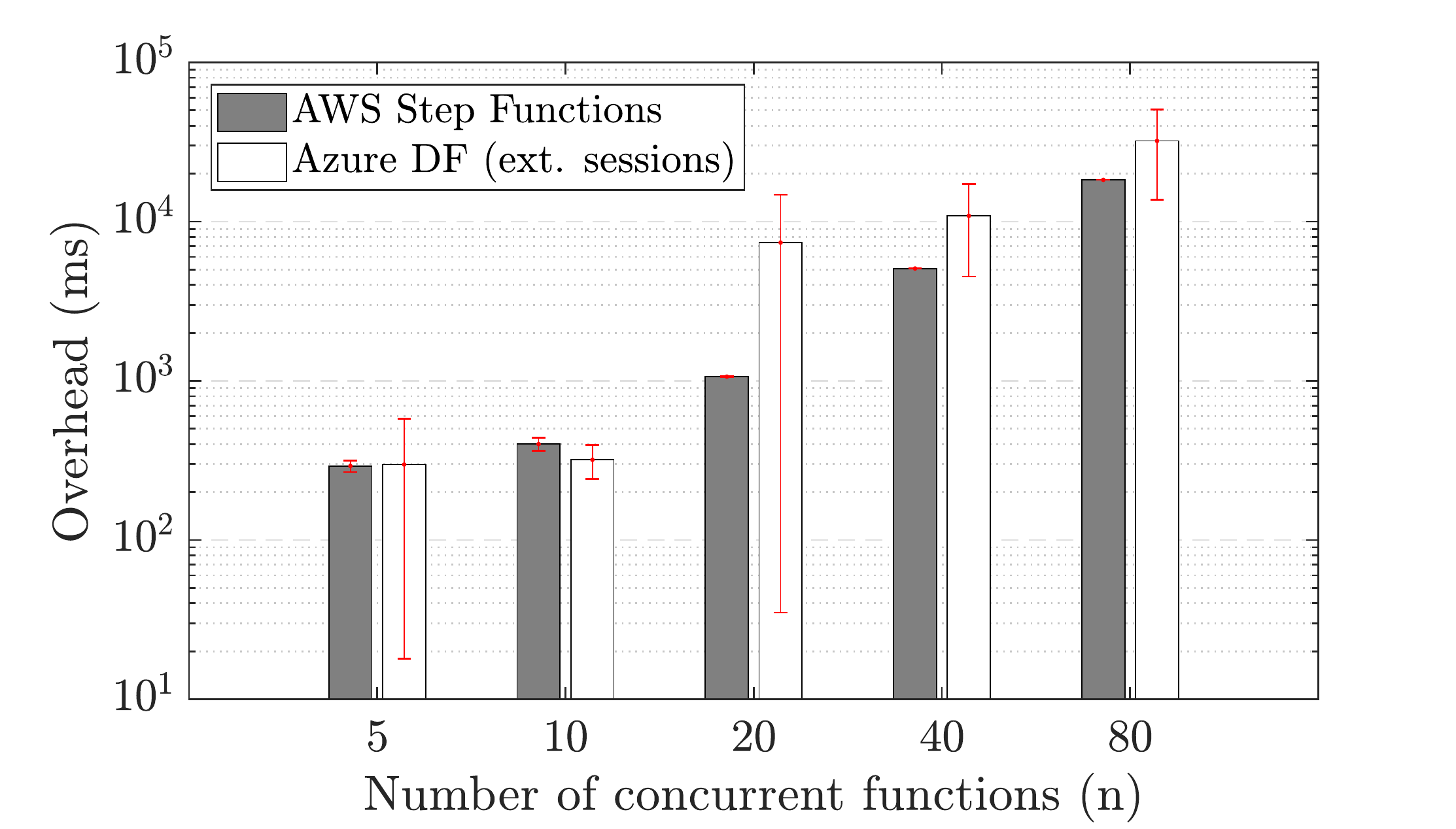}}
  \caption{Parallelism overhead.}
  \label{fig:par}
\end{figure}

\noindent\textbf{Results.} Fig.~\ref{fig:par} depicts the run-time overhead incurred in both platforms when invoking functions in parallel. We can extract two main insights from this figure. First, the overhead grows exponentially with the number of parallel functions $n$. To wit, with $80$ functions, ASF has an average overhead of $18.3$s and ADF of $32.1$s, respectively. Secondly, the results also suggest that ADF exhibits a high variability, whereas overhead on ASF is quite predictable.


\subsection{State management}

First, note that a fair comparison of the overhead attributed to state passing between functions is difficult~in practice. The main reason is that each platform has its own limits on the size of function parameters and results. To wit, ASF has a limit of $32,768$ characters ($32$ KB), whereas Azure Functions provides large message support, allowing parameters and return values of any size (those greater than $60$KB are stored compressed~in Azure Blob Storage). IBM Cloud Functions has a limit of $1$MB for both parameters and results, albeit we empirically found in our tests that this limit is actually of $5$MB.

To measure this overhead, we built a short sequence of $5$ functions, each one receiving a parameter, sleeping for $1$s, and returning the same parameter. We used a payload of $32$KB, as it is the maximum allowed by AWS.

\begin{table}[htbp]
\def\arraystretch{1.5}
\caption{Overhead for a sequence of $5$ actions and a payload of $32$KB.}
\begin{center}
\begin{tabular}{|l|r|r|r|}
\hline
\textbf{Platform} & \multicolumn{2}{c|}{\textbf{Overhead (ms)}} & \textbf{Increase (\%)}  \\
\cline{2-3}
 & \multicolumn{1}{p{1cm}|}{\textbf{Without\newline payload}} & \multicolumn{1}{p{1cm}|}{\textbf{With\newline payload}} & 
\\
\hline \hline
\textit{IBM Sequences} & $49.0$ & $80.8$ & $65\%$ \\ \hline
\textit{IBM Composer} & $175.7$ & $298.4$ & $70\%$ \\ \hline
\textit{AWS Step Functions} & $168.0$ & $287.0$ & $71\%$ \\ \hline
\textit{Azure DF} & $766.2$ & $859.5$ & $12\%$ \\ \hline
\end{tabular}
\label{tab:state}
\end{center}
\end{table}

\medskip
\noindent\textbf{Results.} Table~\ref{tab:state} reports the overhead of this sequence in each platform (baseline) and how increases when state is passed between functions. We observe a clear increase in the overhead in IBM Cloud and AWS, whereas the overhead for Azure slightly increases, remaining high in all cases.

We also found that larger parameters considerably increase the overhead in IBM Cloud. Starting at 500KB (for Composer) and at $1$MB (for Sequences), each state transition adds an extra delay greater than  $10$s. 
On the contrary, the overhead of ADF's large message support  grows linearly with the parameter size, but it is considerably lower than that in IBM Composer.

\section{ST-safe Alternative: Suspend API for Functions}

In this paper, we propose a simple alternative to enable the construction of custom orchestrators. To this end, we propose the following extension:

\medskip
\textit{\texttt{\textbf{Function.suspend(Event)}}}: \textit{This abstraction will move the function to a suspended state linked to a given event \texttt{Event}. The run-time must passivate the current function and stop billing it  until the function is reactivated again by the triggering of the defined \texttt{Event}.}

\medskip
Cloud Providers could, of course, establish time limits to kill suspended states. Since the core run-time is reactive and event-based, suspending a function and triggering its activation with a custom event should be feasible. In this line, the recent introduction of \emph{SQS Custom Event Sources} by Amazon \cite{awstrigger} shows how this triggering could be produced. Passivation and activation of functions could be inspired in previous works on continuations\cite{continuations}.

This simple API would then enable third-party developers to implement their own custom orchestrators that comply with the serverless trilemma. It is obvious that these orchestrators would be \emph{ST-safe}: (I) invocations would not be double-billed (during the suspended state); (II)  substitution principle would be respected, because compositions are normal functions, and (III) composed functions may be black-boxes.

Many programming patterns like async/await, fork/join, fan-out/fan-in may be implemented on top of the \emph{Suspend API}. It is also certain that providing fault-tolerance to state transitions should be then responsibility of the custom orchestrators. The run-time core should only guarantee the recovery  of the last suspended state.

\section{Insights and Future Directions}

First of all, we must consider that IBM Composer and ADF are still experimental projects that could improve in the next months. However, after evaluation and overhead quantification, we are now in position to give some interesting insights, which ensue from our comparison in Table~\ref{tab1}: 

\begin{table}[htbp]
\def\arraystretch{1.5}
\newcolumntype{P}[1]{>{\raggedright\arraybackslash}p{#1}}
\caption{Evaluation Framework.}
\begin{center}
\begin{tabular}{| P{1.8cm} | P{1.8cm} | P{1.8cm} | P{1.8cm} |}
\hline
\textbf{Metrics} &\multicolumn{3}{c|}{\textbf{Systems}} \\
\cline{2-4} 
 & \textbf{\textit{Amazon Step Functions}} & \textbf{\textit{IBM Composer}} & \textbf{\textit{Azure Durable Functions}} \\
\hline \hline
\textit{ST-safe}~\cite{trilemma} & \textit{No} (compositions are not functions) & \textit{Yes} (composition as functions) & \textit{Yes} (composition as functions)  \\ \hline
\textit{Programming model} & DSL (JSON)  & Composition library (Javascript) &\texttt{async/await} (C\#)  \\ \hline
\textit{Reflective API} & \textit{Yes (limited)} & \textit{No} &  \textit{Yes}  \\ \hline
\textit{Parallel execution support} & \textit{Yes} (limited)  & \textit{No} & \textit{Yes} (limited)  \\ \hline
\textit{Software packaging and repositories} & \textit{Yes}  & \textit{Yes} &  \textit{Yes} (no repo) \\ \hline
\textit{Billing model} & $\$0.025$ per $1,000$ state transitions & Orchestrator function execution & Orchestrator function execution + storage costs \\ \hline
\textit{Architecture} & Synchronous client scheduler & Reactive scheduler &  Reactive scheduler \\ 
\hline
\end{tabular}
\label{tab1}
\end{center}
\end{table}

\begin{enumerate}

\item \textit{ASF is the most mature and performant project in the market}: According to the validation, ASF appears to be the most efficient service for both short and long-running orchestrations. IBM is following close for short-running orchestrations. ADF still exhibits significant overhead for all categories. This, of course, can radically change in the future with more stable releases entering the scene.

\item \textit{ADF is the most advanced in terms of programmability. IBM Composer wins in simplicity}:  Coding abstractions in ADF (e.g., async/await,  eternal orchestrator, singleton addressing) are overtly the most advanced, but they are designed for skilled developers. On the contrary, IBM's Composer library is more limited, but also easier to use. ASF programmability is very limited compared~to the other projects.

\item \textit{IBM Composer is designed for short-running sequential orchestrations}:  Unlike ASF or ADF, IBM Composer is not designed to run workflows that last for days or even months. It is now mainly targeting at Web mashups and interactive APIs that require simple workflows. 

\item \textit{None of the existing services is prepared for parallel programming}: Neither ASF nor ADF offer satisfactory concurrency and parallelism for compute-intensive tasks. The overheads are too high. This could change in the future if there is user demand, but until then, external client schedulers will likely be the norm to tap into the massive parallelism of functions.

\item \textit{State size implies costs and overheads}: ASF imposes a stringent limit of $32$KB for state passing, which allows them to offer a clear billing model  and a very stable run-time with predictable overheads. ADF does not set limits on state size. But overheads are still unstable and the final costs are even unknown beforehand. IBM offers $5$MB, but without revealing how the cost of handling state is calculated. Also, its performance declines with state size. In this case, Amazon is the most mature project.

\item \textit{Orchestration should have a cost}: Cloud vendors cannot offer this service for free, because it consumes storage and computational resources.  Only storage and retrieval of state, and fault-tolerance support, incur in derived storage costs. Again, Amazon is the most mature project, and they are the only ones offering a clear billing model. Others will have to follow suit in the next months.

\item \textit{If the cost is high, users will create their own external orchestrators}: We still do not have adoption statistics for these commercial services, but in many cases, users will develop external client schedulers to avoid billing costs. They will not be \emph{ST-safe}, but this is not mandatory for many applications.

\item \textit{Event-based Suspend API for functions may deem run-time orchestrators unnecessary}: Just a simple abstraction like suspending the state of a function until an event is triggered may be the optimal solution. Using this API,~it would be easy to implement a \emph{ST-safe} orchestrator.  All programming abstractions (e.g., sequences, branching, parallel execution) could be built on top of it.

\end{enumerate}



\bibliographystyle{IEEEtran}
\bibliography{IEEEabrv,wosc4}

\end{document}